# Enhanced spectral resolution for broadband coherent anti-Stokes Raman spectroscopy

Faris Sinjab,[1] Kazuki Hashimoto,[2,3] Xuanqiang Zhao,[4] Yu Nagashima,[5] Takuro Ideguchi [1,2,6,*]

1 Institute for Photon Science and Technology, The University of Tokyo, Tokyo 113-0033, Japan

2 Department of Physics, The University of Tokyo, Tokyo, 113-0033, Japan

3 Aeronautical Technology Directorate, Japan Aerospace Exploration Agency, Tokyo 181-0015, Japan

4 Department of Physics, University of California, San Diego, La Jolla, CA 92093, USA

5 Department of Neurology, The University of Tokyo, Tokyo 113-8655, Japan

6 PRESTO, Japan Science and Technology Agency, Saitama 332-0012, Japan

*Corresponding author: ideguchi@ipst.s.u-tokyo.ac.jp

**Abstract**

The spectral resolution of broadband Fourier-transform coherent anti-Stokes Raman spectroscopy is limited by the maximum optical path length difference that can be scanned within a short time in an interferometer. However, alternatives to the Fourier-transform exist which can bypass this limitation with certain assumptions. We apply one such approach to broadband coherent Raman spectroscopy using interferometers with short delay line (low Fourier spectral resolution) and large delay line (high Fourier spectral resolution). With this method, we demonstrate broadband coherent Raman spectroscopy of closely spaced vibrational bands is possible using a short delay line interferometer, with superior spectral resolution to the longer delay line instrument. We discuss how this approach may be particularly useful for more complex Raman spectra, such as those measured from biological samples.

**Introduction**

Coherent Raman spectroscopy is a powerful tool for rapid label-free measurement based on coherent anti-Stokes Raman scattering (CARS) or Stimulated Raman Scattering (SRS), which are particularly powerful for applications in the life sciences [1-3]. Time-domain (TD) coherent Raman spectroscopy using ultrashort pulses for impulsive SRS excitation enables broadband CARS spectroscopy, where the duration of the pulses dictates the shortest molecular vibrational time period (highest frequency, i.e. spectral bandwidth) that can be resolved [4]. Using a setup where pulse pairs with a scanned relative delay excite a sample, the TD-CARS response can be detected at a down-converted timescale set by the delay scanning speed. The spectrum can then be retrieved by Fourier transformation of the TD-CARS interferogram [4]. In this case, the spectral resolution is dependent on the maximum delay achievable between the pulse pairs used for excitation.

Many approaches exist for creating a delay between pulse pairs, including pulse shaping using a spatial light modulator [5], dual frequency-comb excitation [6,7] and mechanical delay scanning using an interferometer [8, 9]. The measurement time of SLM-based techniques is restricted by the refresh rate of the SLM device, typically <1 kHz. Dual-comb CARS techniques can accurately measure a high resolution spectrum in a short time, but the spectral acquisition rate is limited by the recycling time set by the laser repetition rates. Utilizing an interferometer with rapid mirror-based delay scanning can achieve more practical and efficient duty cycles at spectral acquisition rates up to 50 kHz. However, in this case the spectral resolution is restricted due to the properties of the Fourier transform by the maximum scanning time between the pulse pairs, dictated by the optical path difference (OPD) achievable in the scanning arm of the interferometer. Direct delay scanning (e.g. with a resonant scanning mirror) has high throughput, 100% duty cycle, and thus a high photon budget per interferogram delay point, but suffers from low spectral resolution due to the short maximum delay achievable (~1 mm) [8]. Fourier-domain delay scanning (e.g. with a polygonal mirror in a pulse shaper) can achieve similar scan rates with larger delay (>4 mm) and thus an improved spectral resolution, but at a cost to the throughput due to grating-based pulse shaping, lower duty cycle based on the pulse shaper geometry (~50%), and therefore a less efficient utilization of the photon budget [9].

The spectral resolution limitation of the Fourier transform is often treated as an unavoidable fact in time-domain spectroscopy. However, the requirement that a time-domain signal will persist for a sufficient time to permit high spectral resolution is not true for many physical processes of interest, e.g. seismology [10], speech/audio analysis [11], magnetic resonance [12] and optical [13, 14] spectroscopy. For such phenomena, a variety of alternative spectrum estimation techniques have been proposed which can circumvent the limitations of the Fourier transform in some cases [15]. The Raman free induction decay observed with TD-CARS is an example of such a process, where the vibrational excitation decays rapidly with time. As the information content of a real signal containing noise will be concentrated at short delay times, increasing the OPD will often only improve the spectral resolution at the expense of a worse signal-to-noise ratio (SNR).

Here, we investigate the use of an alternative spectrum estimation technique to the fast Fourier transform (FFT) for TD-CARS. The technique models the data as a sum of multiple exponentially decaying sinusoids, from which all appropriate parameters can be estimated and used to plot the spectrum directly. We compare results obtained by applying this procedure to data obtained from a short maximum OPD instrument with the FFT of data from a large maximum OPD instrument, demonstrating that superior spectral resolution is possible with a short delay instrument. Finally, we compare the alternative spectrum estimation procedure with the FFT of a simulated biological dataset with many complex overlapping Raman bands.

**Methods**

An overview of the TD-CARS instrumentation is shown in Fig. 1(c), and is based around a mode-locked Ti:Sapphire laser (Synergy PRO) as a light source, with 75 MHz repetition rate, 790 nm center wavelength, 140 nm FWHM bandwidth, 10 fs pulse width, and 820 mW power output. The beam passes through a variable ND filter wheel and half-wave plate before entering a polarizing beamsplitter (PBSW-10-800, Optosigma) at the entrance of a Michelson interferometer. The Michelson interferometer with short maximum OPD shown in Fig. 1(a) is based on a resonant mirror to scan the delay [8]. The scan arm consists of a quarter-wave plate (AQWP05M-980, Thorlabs), 12 kHz resonant scanning mirror (CRS 12 kHz, Cambridge Technology), curved mirror (CM254-050-P01, Thorlabs), and end mirror. The reference arm consists of another quarter-wave plate, curved mirror, and end mirror. The Michelson interferometer with large maximum OPD shown in Fig. 1(b) is based on a polygonal mirror scanning Fourier-domain delay line, similar to the system described in [9]. The scanning beam passes through a quarter-wave plate into a folded reflective grating-based 4-f pulse shaper (53066BK02-790R, Richardson Gratings, with CM-508-100-P01, Thorlabs) with a polygonal mirror with 18 facets (Lincoln Laser SA34/DT-18) placed in the Fourier plane. The spectrally dispersed beam is reflected off the polygonal scanner and back through the pulse shaper towards a retro-reflecting end mirror. The reference beam passes through a quarter-wave plate, and onto an end mirror placed on a motorized stage. The reference and scan pulses are recombined at the beamsplitter, and pass through a dispersion compensation setup consisting of chirped mirrors (DCMP175, Thorlabs). The beam passes a 700 nm long-pass wavelength filter (FELH0700, Thorlabs) before entering a 50×/0.65 N.A. microscope objective (LCPlanN, Olympus), where the beam is focused onto a sample mounted on a motorized stage (Thorlabs, MLS203-1). The CARS signal collected in transmission mode passes through a short-pass wavelength filter (FESH0700, Thorlabs) and focused onto an APD (PDA10A-EC, Thorlabs). The APD signal is electronically low-pass filtered at 32 MHz (Minicircuits BLP-30+/BLP-50+) before acquisition using a digitizer board at 125 MS/s (AlazarTech, 9440).

The digitized interferogram data is processed in Python (Anaconda distribution) with home-built code based on standard scientific package functions. For each type of interferometric delay line, nonlinear scanning correction is applied to the interferogram data as described previously [8,9]. The conventional FT-CARS spectrum can then be obtained from the absolute value of the fast Fourier transform (FFT) of this interferogram after triangular apodization and zero-padding. Alternatively, the TD-CARS signal could be assumed to be well-approximated by a sum of M decaying sinusoidal functions of the form

$$I(t) = \sum_{k}^{M} A_k e^{j\phi_k} e^{-\alpha_k t} e^{j2\pi f_k t} \qquad (1)$$

where $A_k$, $\phi_k$, $\alpha_k$, $f_k$ are the amplitude, phase, decay constant and frequency respectively for the $k^{th}$ decaying sinusoidal component, and $j = \sqrt{-1}$. A variety of techniques exist which can retrieve information about these parameters for application in spectrum estimation [10,15]. Several such techniques allow the estimation of all four parameters in Eq. (1) with minimal or no initial guesses required for the fitting parameters. One such approach, the Matrix Pencil (MP) method [16], was developed to extract the $4 \times M$ parameters from such data, from which an equivalent spectral profile can be directly constructed using the square root of the Prony energy spectral density formula [15]

$$S(f) = \left| \sum_{k}^{M} A_k e^{j\phi_k} \frac{2\alpha_k}{[\alpha_k^2 + (2\pi[f - f_k])^2]} \right|. \qquad (2)$$

The total number of decaying sinusoidal functions, M, can be selected manually or determined automatically using a noise threshold during the truncated singular value decomposition step of the MP algorithm [16].

**Results**

Fig. 2(a) shows an example of a coherently averaged interferogram of toluene using the polygon mirror FT-CARS system. An OPD of ~3.6 mm, corresponding to roughly 12 ps of group delay is scanned in about 50 µs of measurement time, with the mirror operating at a spectral acquisition rate of 3 kHz, giving a useable measurement duty cycle of approximately 15% [9]. This maximum delay allows CARS spectra with a high spectral resolution to be obtained from FFT processing, as shown in Fig. 2(b). The magnified views of the CARS spectrum in the inset of Fig. 2(b) show toluene Raman bands can be detected across a bandwidth of >1100 cm$^{-1}$, with accurate values for band position determined by automatic peak detection. To test the MP-Prony procedure, the interferogram in Fig. 2(a) is truncated at different measurement times, as indicated by the dashed lines. The resulting CARS spectra from FFT and MP-Prony processing are shown in Fig. 2(c), with the inset highlighting the different around the 787 cm$^{-1}$ band on a semi-logarithmic scale. The spectral resolution of the FFT spectra clearly degrade, while the MP-Prony spectra maintain high spectral accuracy and resolution.

Fig. 3(a) shows a 100-averaged CARS interferogram obtained using the resonant mirror scanning interferometer from a mixture of toluene and benzene. The spectrum obtained by applying an FFT is shown in Fig. 3(b). As expected, due to the shorter delay range, the spectral resolution is worsened in comparison with the polygonal scanner CARS spectrum in Fig. 2(b). Using the full-width half-maximum of the isolated 787 cm$^{-1}$ band shown in the inset of Fig. 3(c) as a metric, the value for the polygonal scanner FT-CARS spectrum is 6.5 cm$^{-1}$ whereas that for of the resonant scanner system is 20.4 cm$^{-1}$. This clearly indicates that the latter system is sub-optimal for resolving closely spaced Raman bands, for example those observed in the Raman spectra of biological samples. This is demonstrated in Fig. 3(c), where for the toluene-benzene mixture, the FFT-derived CARS spectrum cannot resolve the closely spaced aromatic ring stretching modes at 993 cm$^{-1}$ (benzene) and 1004 cm$^{-1}$ (toluene), which appear as a single broad Raman band. The use of the MP-Prony spectrum estimation approach to the interferogram in Fig. 3(a) is shown with the green line in the CARS spectrum in Fig. 3(b) and Fig. 3(c). There is a clear difference in spectral resolution compared with the FFT spectrum using the same interferogram data. For the MP-Prony approach, the FWHM of the 787 cm$^{-1}$ band is now 3.8 cm$^{-1}$, an improvement of roughly 5.3× relative to the FFT spectral resolution, and 1.7× relative to the polygonal scanner with a longer delay line. Furthermore, Fig. 3(c) shows that the closely spaced 993 cm$^{-1}$ and 1004 cm$^{-1}$ bands from benzene and toluene are now clearly distinguishable. A value of M=10 was selected for the spectrum in Fig. 3, which effectively forces the MP method to fit the next most significant data feature as determined by the SVD computation. As seen for the low wavenumber band around 217 cm$^{-1}$, where the single Raman band generates two distinct bands, this may lead to spurious results if M is too high. However, as described in [16], this could be avoided if a noise threshold is used to select an appropriate value for M.

For the measurements of neat liquids shown in Fig. 2 and Fig. 3, the TD-CARS signal will typically extend out to ~10 ps. However, more complex Raman spectra with varied lineshapes will typically not extend this far. Thus TD-CARS instruments with large delay for high spectral resolution will make poor use of the available photon budget, which is an important consideration for both signal generation and avoidance of unnecessary photodamage with unused photons.

Fig. 4(a) shows a simulated example interferogram of a biological dataset, obtained from the inverse FFT of a spontaneous Raman spectrum of a neural stem cell, shown in Fig. 4(b) [17], followed by down-conversion to a resonant scanner instrument timeframe (1 mm maximum OPD at 24 kHz rate). In using a spontaneous Raman spectrum to model the $\chi^{(3)}$-associated TD-CARS spectrum, it is assumed that an excitation pulse would be transform-limited (no higher order dispersion) with sufficiently short duration to adequately probe all vibrational modes within the bandwidth of the model spectrum. The simulated interferogram in Fig. 4(a) clearly decays at a faster rate compared to neat liquids, suggesting that for such samples a resonant scanner instrument with shorter delay would be preferential in terms of allocating the measurement photon budget efficiently. The FFT and MP-Prony spectra from the simulated interferogram are shown in Fig. 4(b), with the model spectrum shown for comparison. The FFT spectrum can resolve some of the larger intensity bands, though the complex spectral profile is lost and the spectral resolution is clearly degraded. For the MP-Prony spectrum however, many of the subtler spectral features which are lost in the FFT spectrum are retained, and in close agreement with the model spectrum. Such spectral features are crucial for the detailed characterization and analysis of bio-samples using chemometric approaches [1]. Therefore, the combination of a short maximum OPD instrument which makes optimal use of the photon budget, combined with alternative spectrum estimation techniques could be a more effective route towards rapid broadband TD-CARS of complex bio-samples, especially when combined with sensitivity enhancement techniques [18-20]. While the MP-Prony spectrum estimation approach is based on a physically-relatable model of the time-domain data, other more contemporary approaches such as compressive spectral sensing may further improve the spectrum estimation accuracy and computational performance [21].

**Conclusion**

In summary, we have demonstrated an alternative spectrum estimation approach to the FFT for enhancing the spectral resolution of TD-CARS spectroscopy when using a short delay line for measurement of closely spaced spectral lines. Using simulated data, we suggest that this approach may be especially useful for efficient measurement of high-resolution CARS spectra of complex bio-samples with a range of closely spaced Raman bands of varying linewidth.

**Funding**

F.S. and X. Z. thank financial support by JSPS and UCEAP program, respectively. This work was financially supported by the New Energy and Industrial Technology Development Organization (NEDO) project "Development of advanced laser processing with intelligence based on high-brightness and high-efficiency laser technologies", JST PRESTO (JPMJPR17G2), and JSPS KAKENHI (17H04852, 17K19071, 18F18012).

**Acknowledgments**

We thank Prof. J. Yumoto and Prof. M. Gonokami for use of their equipment.

**Figures**

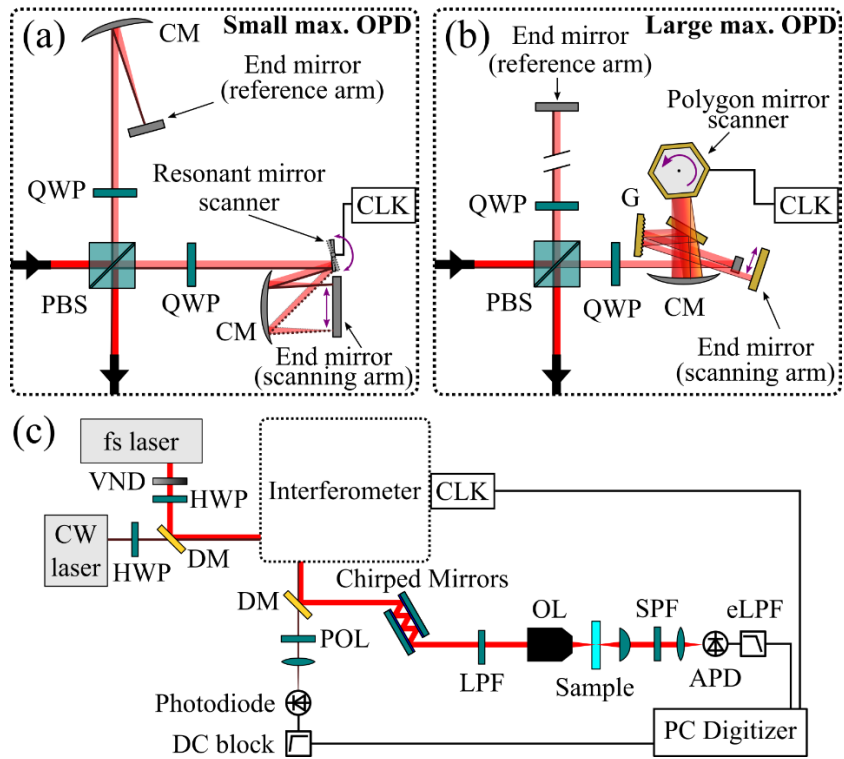

Fig. 1. Broadband CARS using two different interferometric delay lines. (a) Short OPD delay line based on a Michelson interferometer with a resonant mirror scanning arm [8]. (b) Long OPD Fourier-domain delay line based on a Michelson interferometer with a 4-f pulse shaper and polygon mirror scanner [9]. (c) CARS instrument schematic. VND: Variable Neutral Density filter, HWP: Half-Wave Plate, DM: Dichroic Mirror, PBS: Polarizing BeamSplitter, QWP: Quarter-Wave Plate, CM: Curved Mirror, G: Diffraction Grating, POL: Linear Polarizer, LPF: Long-Pass wavelength Filter, OL: Objective Lens, SPF: Short-Pass wavelength Filter, APD: Avalanche Photodiode Detector, eLPF: electronic Low-Pass Frequency Filter. CLK: scanning mirror clock signal.

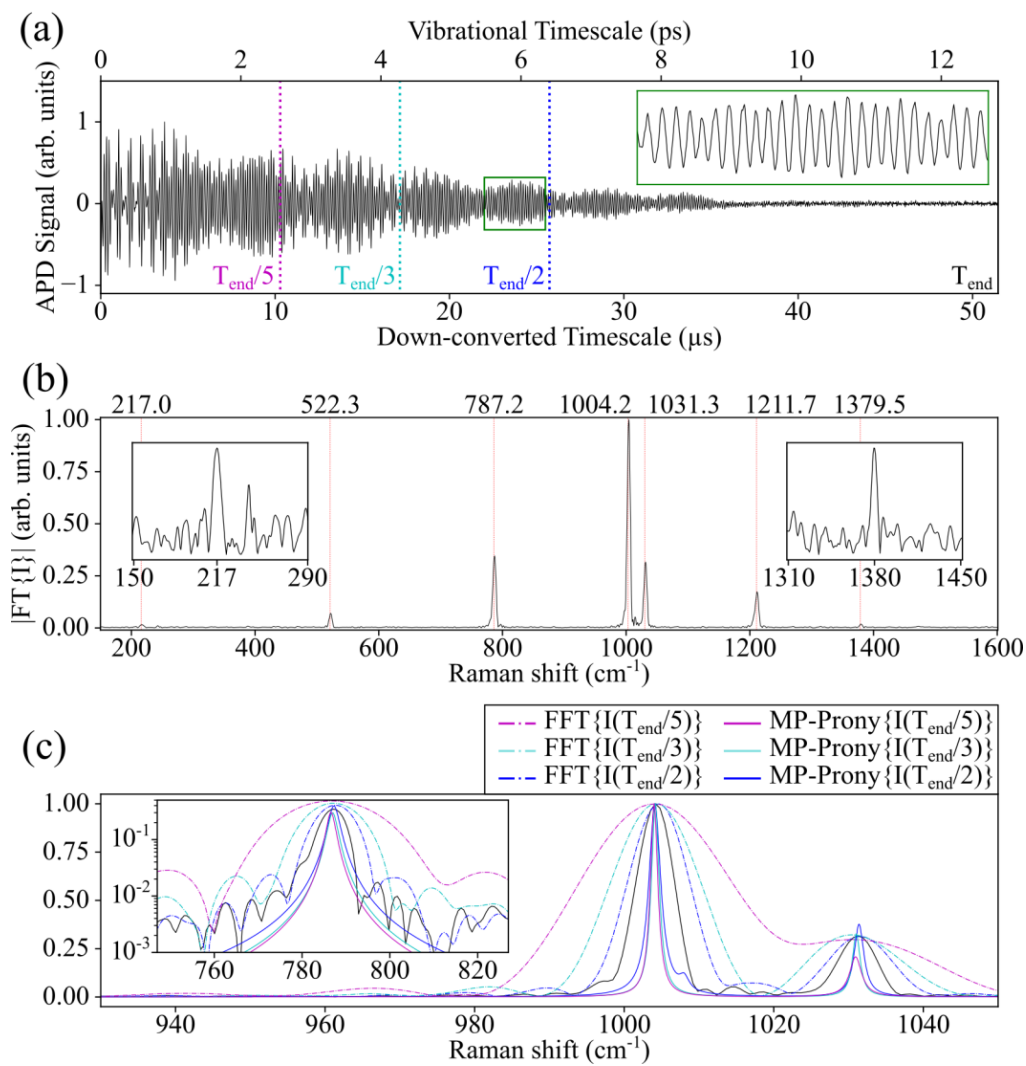

Fig. 2. Broadband CARS using an interferometric delay line with large maximum OPD. (a) 1000-averaged CARS interferogram of toluene. (b) CARS spectrum obtained from the FFT of the interferogram in (a), with band positions from automated peak fitting shown. (c) Comparison of FFT and MP-Prony spectrum estimation (M=10) applied to the interferogram in (a) after truncation to the times illustrated in (a).

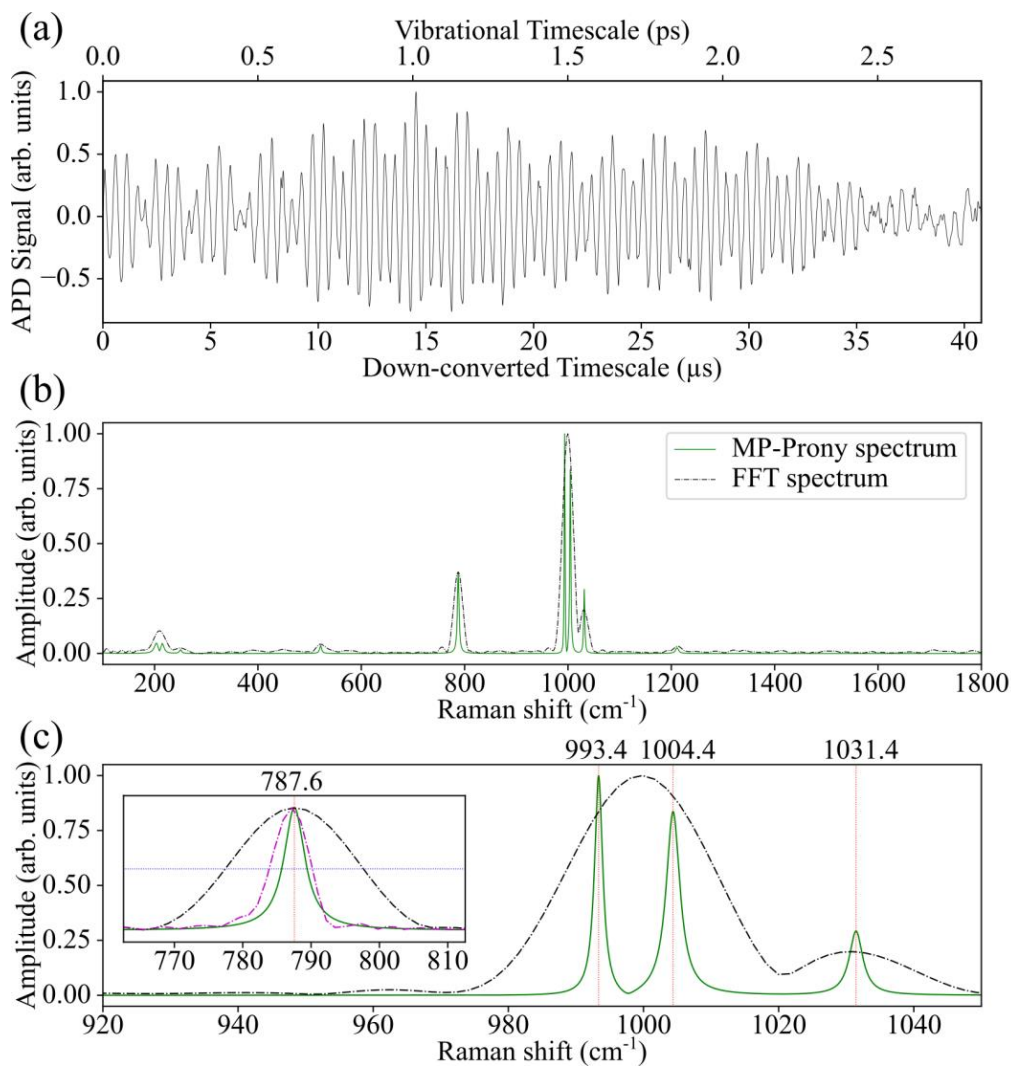

Fig. 3. Broadband CARS of a mixed sample utilizing a short OPD delay line interferometer. (a) 100-averaged CARS interferogram of toluene-benzene 2:1 mixture. (b) CARS spectra obtained by applying either FFT or MP-Prony (M=10) techniques to the interferogram in (a). (c) Magnified view of the aromatic ring stretching region of the CARS spectrum, and inset showing the 787 cm$^{-1}$ toluene band with the polygonal mirror FFT spectrum from Fig. 2(b) also shown (magenta dashed line). The blue line is a guide for the half-maximum position.

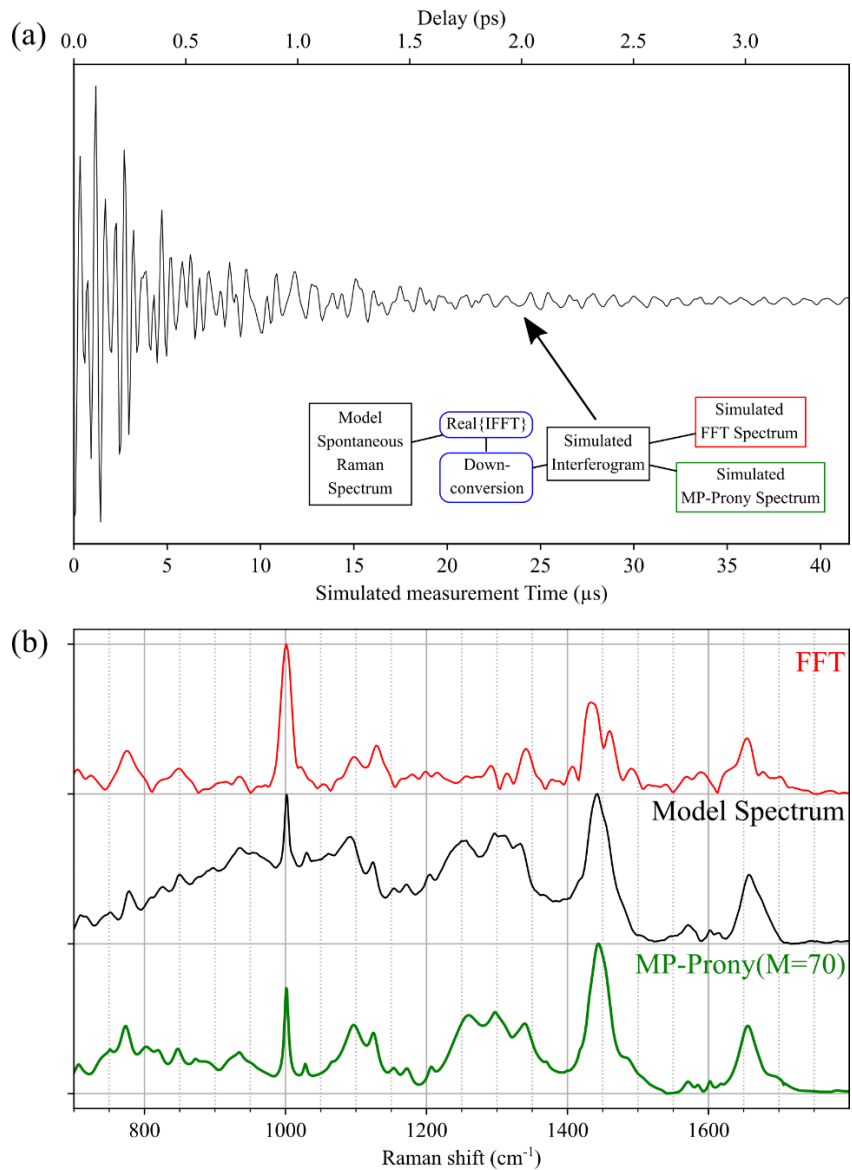

Fig. 4. Simulated TD-CARS bio-sample dataset for a short maximum OPD delay line interferometer. (a) Simulated interferogram generated from the model spontaneous Raman spectrum shown in (b) by the procedure shown in the inset flow diagram. (b) FFT and MP-Prony spectrum estimation using the simulated interferogram in (a), with the model input spectrum for reference.